\documentclass{article}

    \usepackage[preprint]{neurips_2019}

\usepackage[utf8]{inputenc} 
\usepackage[T1]{fontenc}    
\usepackage{hyperref}       
\usepackage{url}            
\usepackage{booktabs}       
\usepackage{amsfonts}       
\usepackage{nicefrac}       
\usepackage{microtype}      

\usepackage{caption}
\usepackage{subcaption}
\usepackage{float, graphicx, wrapfig}
\usepackage{booktabs, tabularx, multirow, tablefootnote}

\setcitestyle{author year,open={(},close={)}}

\title{Graph Representation learning for Audio \& Music genre Classification}

\author{%
  Shubham Dokania\textsuperscript{*} \\
  Mercedes-Benz R\&D India\\
  Bangalore, India\\
  \texttt{shubham.dokania@daimler.com} \\
  \And
  Vasudev Singh\thanks{Equal Contribution} \\
  Mercedes-Benz R\&D India \\
  Bangalore, India \\
  \texttt{vasudev.singh@daimler.com}} 

\begin{document}

\maketitle

\begin{abstract}
Music genre is arguably one of the most important and discriminative information for music and audio content. Visual representation based approaches have been explored on spectrograms for music genre classification. However, lack of quality data and augmentation techniques makes it difficult to employ deep learning techniques successfully. We discuss the application of graph neural networks on such task due to their strong inductive bias, and show that combination of CNN and GNN is able to achieve state-of-the-art results on GTZAN, and AudioSet (Imbalanced Music) datasets. We also discuss the role of Siamese Neural Networks as an analogous to GNN for learning edge similarity weights. Furthermore, we also perform visual analysis to understand the field-of-view of our model into the spectrogram based on genre labels.
\end{abstract}

\section{Introduction}
The advent of large collection of musical data has given rise to the need for aggregation and analysis on a diversity of tasks such as recommendation, classification based on artists (\cite{Berenzweig2002}), generation of music (\cite{Briot2017}). Recent research on Music Information Retrieval (MIR) (\cite{Typke2005}) focuses on task specific methods to address the representation of data. Traditional techniques usually rely on visual features such as MFCC  (\cite{Logan2000}), as input to classifiers. Several approaches try to model the sequential nature of audio by leveraging recurrent architectures such as LSTM (\cite{Tang2018,Dai2016}), whereas there exist approaches which take the combination of sequential as well as spatial information for this task (\cite{Senac2017,Choi2017}). (\cite{Liu2019}) proposed a broadcasting mechanism to better represent audio features for processing in CNNs and reported state-of-the-art results on GTZAN dataset.

Graph Neural networks, introduced in (\cite{Gori2005,Scarselli2008}) are based on processes involving local operations on graph and graph nodes. Graph neural networks exhibit strong relational inductive bias (\cite{Battaglia2018}), and hence have been shown to perform very well on semi-supervised tasks, and in modelling of entities  which do not display a natural order. (\cite{Kipf2016}) provided an efficient variant of the GNN, called Graph Convolutional Network (GCN). More recent variations of the GCN have achieved commendable results on problems  including but not limited to natural language processing (\cite{Yao2019,Beck2018,Marcheggiani2018}), computer vision (\cite{Li2017,Qi2017,Gu2018}), applied chemistry and Biology (\cite{Fout2017,Do2019,Duvenaud2015}), and social networks (\cite{Qiu2018,Liu2019}).

In this work, we target the problem of music genre classification in a data-efficient manner, while trying to maximise the representation capability of the inferred features. We divide our approach in two parts: A robust feature representation method based on similarities between fine details observed in the musical data, and a method to explore arbitrary relationships that may exist between multiple audio data files. Our approach builds over the established literature about graph neural networks as a natural alias for convolution over non-euclidean relationships (\cite{Battaglia2018,Kipf2016}).

\begin{figure}[t!]
    \centering
    \includegraphics[width=\linewidth, height=0.420\linewidth]{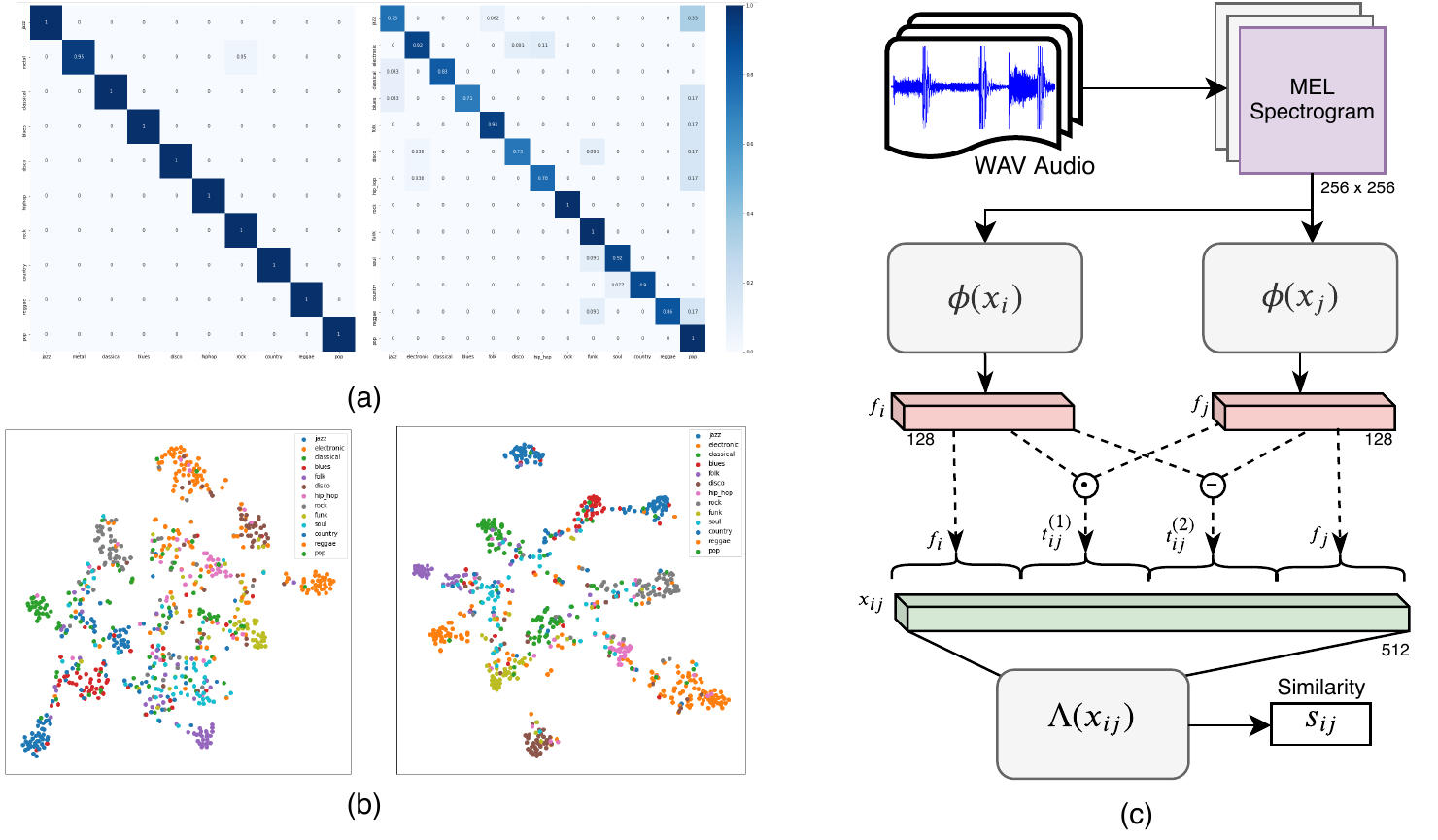}
    \caption{(a) Confusion Matrix for GTZAN and AudioSet datasets from GrahAM (b) t-SNE plots for 50\% of the complex AudioSet data from (i) Siamese Network  and (ii) GrahAM. (c) Illustration of the Siamese Training flow. }
    \label{fig:siamese}
\end{figure}

\section{Our Approach}

\begin{figure}
\centering
    \includegraphics[width=\textwidth, height=5.2cm]{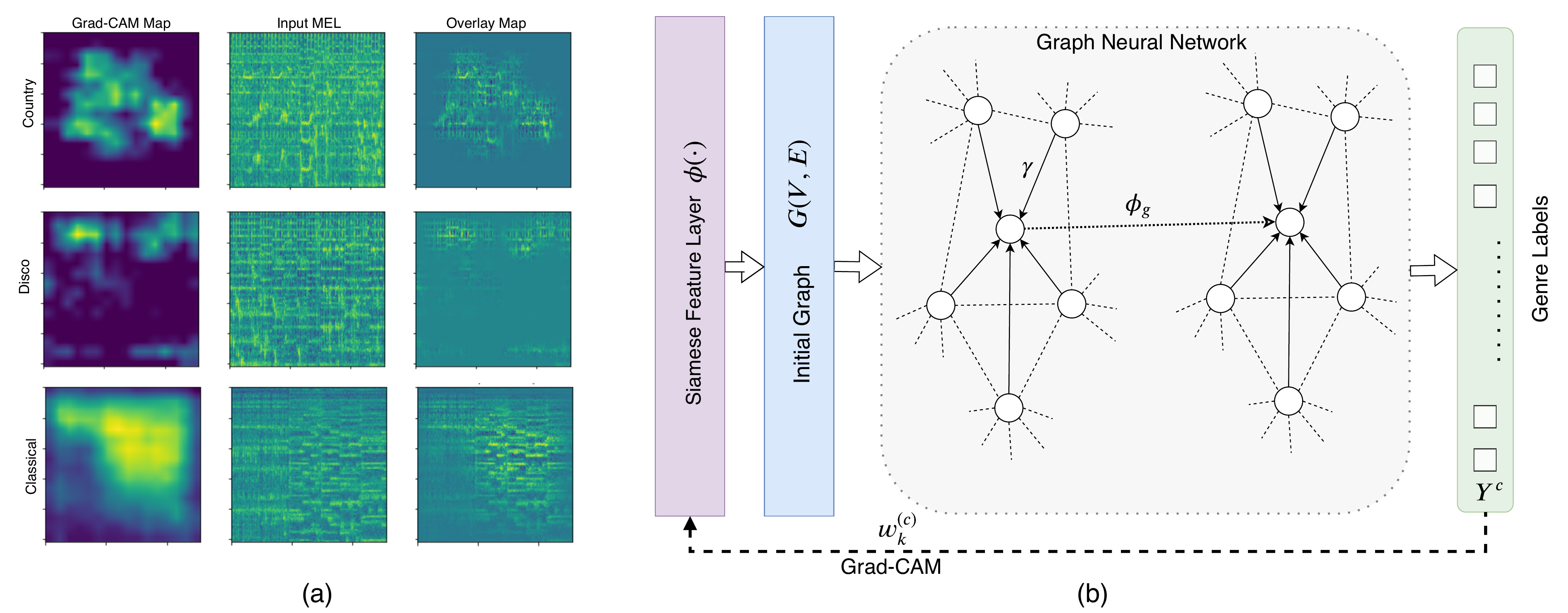}
    \caption{(a) Visualization for the heatmaps generated from Grad-CAM method over example genere predictions from GrahAM. (b) Illustration of the Graph Network processing pipeline with reference to gradient taken from intermediate layers.}
    \label{fig:graph}
\end{figure}

We model the problem of audio classification as node-level classification task for graph neural network, where each node is a compressed vector representation of a MEL spectrogram of the respective audio data. The following section outlines our proposed approach for feature extraction via Siamese module, and discovering structural relations using Graph Neural Network. We also discuss techniques to visually analyse and understand the model. The approach is illustrated in figure \ref{fig:graph}.

\subsection{Siamese Neural Network}

Siamese Neural Networks, first introduced in (\cite{Bromley1994}) to solve signature verification tasks in the form of image matching problem. A Siamese network consists of twin parallel networks which take in two inputs. The parameters between the twin feature generation branches of the network are shared, which guarantees that two very similar images could not possibly be mapped to very different feature vectors in the feature space because each network computes the same function and the top conjoining layers guarantees that songs with different genres are not mapped together. The discriminative nature of the network makes it a good choice to learn robust representations between different music genres which are usually very similar and could pose some difficulty in being learned by other commonly used techniques.

The Siamese network could also be seen in the form of a Graph Neural Network, as also noted by (\cite{Garcia2017}). For a set of input nodes $(x_i, e_{ij}, x_j) \in G(V_{mel}, E_{mel})$, ($G$ being the input graph for MEL data), The existence of edges can be seen as the pairs of spectrograms $(x_i, x_j)$ which are being considered for comparison. We use Xception Net (\cite{Chollet2017} as the backbone for generating the features (written as function $\phi$). Writing the top conjoining layer as $\Lambda$, we can write the overall Siamese similarity method as $s_{ij} = \Lambda (\phi(x_i), \phi(x_j))$. Our Siamese feature branch ($\phi$) takes in an input of the audio in wav vector format, and generates an output of 6x6x128 feature maps. These are then passed through a Global Average Pooling layer to reduce it down to a feature vector of size 128. This process is illustrated in figure \ref{fig:siamese}(c). The lower branch of the siamese network takes the feature vectors from the conjoining layer and applies minor modifications to capture different variations in the available information.

\begin{equation}
x^{(1)}_{ij} = \mathbf{CONCAT}(f_i, f_j) ; \textbf{   }
t^{(1)}_{ij} = (f_i - f_j)^2 \ ; \ t^{(2)}_{ij} = f_i \odot f_j
\end{equation}

\begin{equation}
x^{(2)}_{ij} =\mathbf{CONCAT}(t^{(1)}_{ij}, t^{(2)}_{ij}); \textbf{   }
x_{ij} = \mathbf{CONCAT}(x^{(1)}_{ij}, x^{(2)}_{ij})
\end{equation}

 The above equations summarise the process we use to take combinations such as square distance in $t_{ij}^{(1)}$, dot product in $t_{ij}^{(2)}$, and the overall culmination of the features in $x_{ij}$.

\subsection{Graph Neural Network}

The strong relational inductive bias in graph neural networks makes it a good choice to model arbitrary relationship structure among the components present in the data \cite{Battaglia2018}. Using local aggregation and processing operations on vertices of the graph, it is possible to obtain more expressive representation of the encoded information. In general, graph neural networks are similar to applying convolution neural networks and provide a way to process non-euclidean data.The local operations enable expanding the field of view in the graph network so as to facilitate the maximal flow of information among related nodes and reduce the impact of noise from unrelated nodes. A generic form of defining message passing in a graph network for a given graph $G(V, E)$ can be written as defined by \cite{Fey2019}:

\begin{equation}
x_i^{(k)} = \gamma^{(k)} (x_i^{(k-1)}, F_{j \in \mathcal{N}(i)}(x_i^{(k-1)}, x_j^{k-1}, e_{ij}))
\end{equation}

Where, $x_i, x_j \in V, e_{ij} \in E$ are the nodes and vertices of the graph $G$. The aggregation operation $F$ is applied on all neighbours $\mathcal{N}(i)$ of $i$, to be passed on to a non-linear transformation function $\gamma$ for a given layer $k$. The Graph neural network architecture we considered for experiment is a variant of the Graph Convolution Network (GCN) similar to the Graph Edge Convolution Operation (\cite{Simonovsky2017}). In context of the current discussion, each node $x_i$ in the graph initially represents the features generated from the Siamese module as a 128-length vector. Each node in the graph is assumed to be connected to every other node. A simple comparison between the features as part of the local aggregation process is shown to be able to reduce the impact of noise arising from dissimilar features. In general GNNs, the network depth is chosen to be order of graph diameter, so as to propagate node information while increasing the receptive field, but in our case we can limit the number of layers to a relatively small depth since all nodes are able to pass information internally. We model the message passing between the nodes $x_i, x_j \in G(V, E)$ as following:

\begin{equation}
x_i^{(k)} = \phi_{g}(x_i^{(k-1)}, \gamma(x_i^{(k-1)}, x_j^{(k-1)}))
\label{eq:edge_gcn}
\end{equation}

Where we model the function $\phi_{g}$ and $\gamma$ as feed forward neural networks with Rectified Linear Unit as non-linearity. More specifically, our choice of $\gamma$ takes input $(x_i^{(k-1)} - x_j^{(k-1)})$ to model the similarity/dissimilarity between the two nodes, and the functions $\gamma$ and $\phi_{g}$ control the flow of information. Unlike (\cite{Garcia2017}), we do not consider $abs(\cdot)$ or mean squared over the term  $(x_i^{(k-1)} - x_j^{(k-1)})$ to avoid symmetry in the function. Symmetric information flow is not desirable in our case since the information from $\gamma(x_i^{(k-1)}, x_j^{(k-1)})$ may be useful for the current $x_i^{(k-1)}$ but may not be useful for $x_j^{(k-1)}$. The iterative message passing operation in a graph neural network affects the aggregation and forward flow of information by a significant margin, thus a wise choice of said function remains crucial. Since we consider all nodes to be connected to each other, importance should be put on understanding and approximating the relationship between two nodes $x_i$ and $x_j$, such that the aggregated outcome from the local operator generates constructive features for mutually important nodes, and discard the nodes pairs which do not affect each other. These operations are handled very effectively using the PyTorch-Geometric package (\cite{Fey2019}). As a result of this construction, the network is able to easily adapt to new input nodes since it is assumed to be connected to every other node as well.

\subsection{Grad-CAM Analysis}
Weakly supervised localisation has been widely popular in literature (\cite{Shi2013,Selvaraju2017}) for tasks related to image data. More specifically, approaches to gather information about where a model looks in the image while making a predictions is well studied and several approaches exist which are able to use trained classification models to provide such information (\cite{Chattopadhay2018}). Albeit our data does not strictly lies in the spatial domain where texture and local information collectively construct visually separable objects in an image, our methods rely on the textural information available in the spectrogram data. We show, based on tracking of backward gradient flow and it's magnitude, that our network is able to learn distinctions in the data in quite early stages. Our approach for this analysis is based on the work done by (\cite{Chattopadhay2018}) for providing explainable understanding of the proposed approach. The authors argue that based on the magnitude of gradients for each class score $Y^c$ with respect to the feature maps $A^{k}$ from a layer $k$, a weight can be assigned to each feature map in the following way:

\begin{equation}
w_k^c = \sum_i \sum_j \alpha_{ij}^{kc} \cdot relu (\frac{\partial Y^c}{\partial A^k_{ij}}); \textbf{   } \alpha_{ij}^{kc} = \left\{
\begin{array}{ll}
      \frac{1}{\sum_{l,m} \frac{\partial y^c}{\partial A^k_{lm}}} & if \ \frac{\partial y^c}{\partial A^k_{ij}} = 1 \\
      0 & otherwise
\end{array} 
\right.
\end{equation}

Where, $relu$ is the Rectified Linear Unit activation, and $\alpha_{ij}^{kc}$'s are weighting co-efficients for the pixel-wise gradients for class $c$ and convolutional feature map $A^k$. We refer the reader to (\cite{Chattopadhay2018}) for a more intuitive understanding.

\section{Experiments And Results}


\subsection{Experimental Setup}

For training the model, we use the Adam optimizer with a learning rate of 3e-04 for both the siamese and graph networks. We selectively prune the data combinations in the siamese network since the number of positive examples and negative examples are quadratically unbalanced as follows:

\begin{equation}
C_{total} = {n \choose 2} = \frac{(n)(n-1)}{2}; \textbf{    }
C_{pos} = {k \choose 2}; C_{neg} = {n-k \choose 2}
\end{equation}

Where, $k$ is considered to be the number of labelled (training) samples per class, $C_{total/pos/neg}$ are the number of choices for total, positive and negative cases respectively. Since the distribution could be highly skewed, we select only a fraction of examples from the positive and negative cases to maintain a balance of approximately $1:1$ between the two sets. The Siamese network is then trained for 20 epochs with a batch size of 64 on a Volta V100 GPU, which takes approximately 3.5 hours for the training set. The Graph Network is trained for 1000 epochs for the GTZAN dataset and takes approximately 20 minutes on the V100 GPU. It is important to note that image based augmentation techniques are not applicable in the provided scenario since the location and intensity of each pixel corresponds to specific frequency and sound information in the original audio signal, and any direct transformation may cause the loss of actual meaning.

\subsection{Datasets}

The two main datasets we have considered for our experiments are the GTZAN (\cite{Tzanetakis2002}) and the Unbalanced section of music data from AudioSet (\cite{Gemmeke2017}). GTZAN dataset has been widely used in many studies with the aim of music genre classification, consists of 1000 audio samples, 30-seconds each. Although GTZAN data provides a clean distribution of audio samples, it fails to capture more recent trends in the domain of music (especially, classes such as electronic don't exist in GTZAN). AudioSet consists of an expanding ontology of 632 audio event classes and 2,084,320 human-labelled 10-second sound clips drawn from YouTube videos. It is important to note that image-based augmentation techniques are not applicable in spectrograms since the location and intensity of each pixel corresponds to specific audio-frequency information, and any direct transformation may cause the loss of actual information.

\begin{table}[h!]
    \begin{subtable}{0.48\linewidth}\centering
    {\begin{tabular}{|c | c | c |}
         \hline
         \textbf{Method} & \textbf{Pre-} & \textbf{Acc} \\
            & \textbf{Process} &  (\%) \\
         \hline
         
         Bottom-Up & MEL  & 93.9 \\
         Broadcast  & Spectrogram &  \\
         
         \hline
         Multi-DNN & MFCC & 93.4 \\
         \hline
         CVAF & MEL &  \\
           & Spectrogram, &  90.9\\
              & SSD, etc.        &      \\
        \hline
        GrahAM  & MEL & \textbf{99.5} \\
            & Spectrogram &  \\
         
         \hline
    \end{tabular}}
    \captionof{table}{ }
    \end{subtable}
    \begin{subtable}{0.5\linewidth}\centering
    {\begin{tabular}{|c | c | c | c|}
         \hline
         \textbf{Method} & \textbf{Data} & \textbf{GTZAN} & \textbf{AudioSet} \\
           & \textbf{Split}  & Accuracy & Accuracy \\
         \hline
         
         GraphNN    &   30  &   \textbf{66.4}   &   30.4  \\
         RF      &   30  &   61.0            &   30.0             \\
         2-layerNN         &   30  &   60.0            &   \textbf{31.0}              \\
         \hline
         GraphNN    &   50  &   \textbf{84.9}  &   \textbf{57.9}  \\
         RF      &   50  &   80.5      &   55.0             \\
         2-layerNN         &   50  &   81.0       &   56.0              \\
         \hline
         GraphNN    &   100  &   \textbf{99.5}  &   \textbf{91.3}  \\
         RF      &   100  &   98.0      &   86.0             \\
         2-layerNN         &   100  &   98.5       &   87.0             \\
         \hline
    \end{tabular}}
    \captionof{table}{ }
    \end{subtable}
    \caption{Results showing our method (GrahAM) in terms of accuracy against other methods for (a) GTZAN and (b) Data splits of both GTZAN and AudioSet.}
    \label{tabtab}
\end{table}

\subsection{Quantitative and Visual Analysis}

We discuss about our results on GTZAN as compared to the accuracy values reported in (\cite{Liu2019}) for the full dataset when trained on a 70:30 (train:test) split. As shown in Table \ref{tabtab}(a), Our appraoch, GrahAM (Graph NN with Siamese) is showing significant improvement (99.5\%) to establish state-of-the-art results. We also divide the labelled dataset (for both GTZAN and AudioSet) into smaller labelled fractions (shown in Table \ref{tabtab}(b)). In each of the splits, GNN approach is showing highest performance, compared to that of Random Forest (RF) and a Neural Network classifier, as displayed in Table \ref{tabtab}(b). AudioSet data distribution is relatively more complex than GTZAN and contains several confusing examples due to the trivial similarity between a number of musical genres such as rock and electronic, jazz and pop, etc. This is more clearly visible in the confusion matrices provided in Figure \ref{fig:siamese}(a).

In figure \ref{fig:graph}(a), the three columns represent (i) The heatmap generated from Grad-CAM for the respective class, (ii) the input MEL-Spectrogram, and (iii) the overlay section of the MEL Spectrogram based on the heatmap information. The third column is crucial for our analysis as the emergent pattern signifies the regions of importance in the frequency-time domain for the texture of the tones and chords to be seen as a particular genre by the model. The gradient analysis method almost always reveals the same pattern of importance which constitutes a given genre.The position of high intensity is not as important as much as the texture in the region of importance as we have noticed across our experiments. To justify our claims further, we also display a t-SNE visualization of the features generated from both Siamese and GrahAM in figure \ref{fig:siamese}(b). We notice that the siamese network is able to effectively capture similarity relations, however, some degree of confusion still remains. After a fine-grained processing through GrahAM, the final points are clustered (isolated) more densely. These visualizations establish the representation capability of our approach which exploited the relationship between similar/dissimilar nodes of the graph.









\bibliography{ref}

\end{document}